\documentclass[preprint,3p,times,twocolumn,authoryear]{elsarticle}

\usepackage{graphics}
\usepackage{subfigure}

\usepackage{amssymb}
\usepackage{multirow}

\biboptions{comma,round}

\journal{JOURNAL OF NETWORK AND COMPUTER APPLICATIONS}

\begin{document}

\begin{frontmatter}



\title{Comparative study of High-speed Linux TCP Variants over High-BDP Networks}


\author[lable1]{Mohamed A. Alrshah$^{1,}$}
\author[lable1]{Mohamed Othman$^{2,}$}
\author[lable2]{Borhanuddin Ali}
\author[lable1]{Zurina Mohd Hanapi}
\address[lable1]{Department of Communication Technology and Network, Universiti Putra Malaysia, 43400 UPM, Serdang, Selangor D.E., Malaysia}
\address[lable2]{Department of Computer and Communication Systems Engineering, Universiti Putra Malaysia, 43400 UPM, Serdang, Selangor D.E, Malaysia.}

\begin{abstract}
Transmission Control Protocol (TCP) has been profusely used by most of internet applications. Since 1970s, several TCP variants have been developed in order to cope with the fast increasing of network capacities especially in high Bandwidth Delay Product (high-BDP) networks.  In these TCP variants, several approaches have been used, some of these approaches have the ability to estimate available bandwidths and some react based on network loss and/or delay changes. This variety of the used approaches arises many consequent problems with different levels of dependability and accuracy. Indeed, a particular TCP variant which is proper for wireless networks, may not fit for high-BDP wired networks and vice versa. Therefore, it is necessary to conduct a comparison between the high-speed TCP variants that have a high level of importance especially after the fast growth of networks bandwidths. In this paper, high-speed TCP variants, that are implemented in Linux and available for research, have been evaluated using NS2 network simulator. This performance evaluation presents the advantages and disadvantages of these TCP variants in terms of throughput, loss-ratio and fairness over high-BDP networks. The results reveal that, CUBIC and YeAH overcome the other high-speed TCP variants in different cases of buffer size. However, they still require more improvement to extend their ability to fully utilize the high-speed bandwidths, especially when the applied buffer is $near-zero$ or less than the BDP of the link.

\end{abstract}

\end{frontmatter}

\footnotetext[1]{Corresponding authors: 
\\E-mail addresses: mohamed.asnd@gmail.com (Mohamed Alrshah),
\\mothman@upm.edu.my (Mohamed Othman).
}

\footnotetext[2]{The author is an associate researcher at the Computational Science and Mathematical Physics Lab, Institute of Mathematical Science, Universiti Putra Malaysia.}


\section{Introduction}
\label{Intro}
Transmission Control Protocol (TCP) is commonly used by most of Internet applications and becomes one of the two original components of the Internet protocol suite, complementing the Internet Protocol (IP), thus the entire suite is known as TCP/IP. TCP provides stable and reliable delivery of data packets without relying on any explicit feedback from the underlying network. However, it relies only on the two ends of the connection which are sender and receiver. That is why TCP is known as end-to-end or host-to-host protocol. In the last couple of years, TCP is profusely used by major Internet applications such as file transfer, email, World-Wide-Web and remote administration.

The first idea of TCP had been presented by \citet{Vinton1974}. Thereafter, TCP has been implemented in several operating systems and examined in real environment. With the advancement in network technology, TCP faced many new scenarios and problems, such as network congestion, under utilization of bandwidth, unfair share, unnecessary retransmission, out of order delivery, non-congestion loss. All of these problems encouraged researchers to review the behavior of TCP. In order to solve these problems, many TCP variants have been developed. Each TCP variant has been designed to solve certain problems, some try to survive over a very slow and congested connections, and some try to achieve higher throughput to fully utilize the high-speed bandwidths, while some try to be more fair. In fact, they are mostly different from each other so that categorizes them into high-speed, wireless, satellite and low priority. Indeed, a particular TCP variant which is proper for wireless networks, may not fit for high-BDP wired networks and vice versa.

Therefore, it is necessary to conduct a comparison between TCP variants that are designed for high-speed networks to show the advantages and disadvantages of each TCP variant. In this paper, Scalable TCP, HS-TCP, BIC, H-TCP, CUBIC, TCP Africa, TCP Compound, TCP Fusion, NewReno, TCP illinois and YeAH have been evaluated using NS2 network simulator. This performance evaluation presents the advantages and disadvantages of the compared TCP variants and shows the differences between them in terms of throughput, loss-ratio and fairness over high-BDP networks. As well as, it presents and explains the behaviors of the compared TCP variants, shows the impacts of the used approaches, and arranges the thoughts. Thus, this paper may help the researchers to improve the performance of the existing TCP variants by cutting down the effort of comparing the existing protocols in order to improve it to fit the new generation of the networks.

The rest of this paper is organized as follows: Section \ref{Mot} presents the motivations behind this work, challenges and previous works. While, Section \ref{PE} presents the performance evaluation of high-speed TCP variants and explains the experiments' setup, network topology, performance metrics, results and discussion. Finally, Section \ref{Conc} concludes the paper with some final comments.

\section{Motivations, Challenges and Previous Works}
\label{Mot}
The rapid growth of network technologies reduces the ability of TCP to fully utilize the resources of these networks. Due to this problem of under-utilization of network resources, many high-speed TCP variants that aim to increase the utilization of these resources have been exist. These increase of TCP aggressiveness, in order to fully utilize the high-speed bandwidths, arises the severe problem of burst loss \citep{Ha2008}. In addition to that, the  variety of these TCP protocols leads to some questions that need to be addressed: Which TCP variant seems to be the best for high-speed networks? Are the current TCP variants sufficient to fully utilize the high-speed bandwidths? In order to answer these questions, a comparative study of high-speed TCP variants is required. Such comparison or performance evaluation addresses the points of TCP weaknesses and consequently supports the process of enhancing TCP performance.

\begin{figure}[hbtp]
\includegraphics[width=\linewidth]{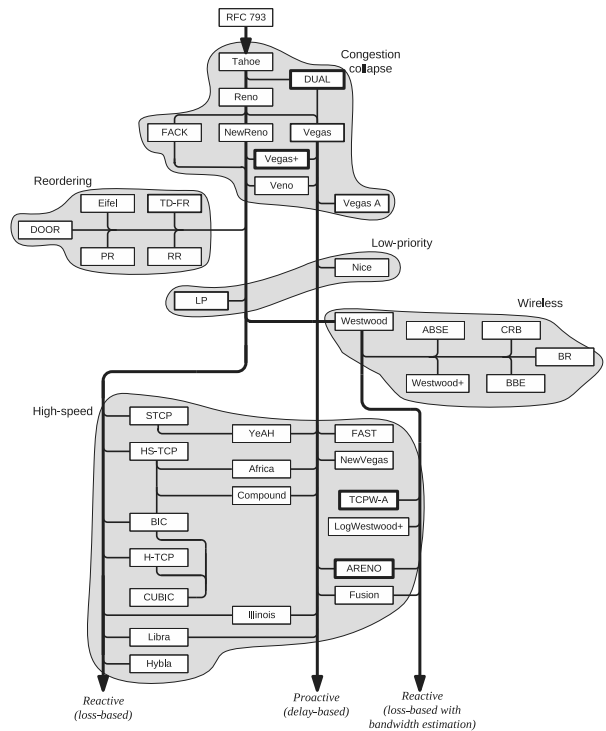}
\caption{The classification and evolution of variants of TCP congestion control \citep{Afanasyev2010}.}
\label{taxonomy}
\end{figure}

Nowadays, TCP is struggling to deal with different network environments such as wireless or lossy networks, high-speed networks and highly congested networks. Each type of these networks has its own problems and limitations that are different from one to another networks. Consequently, there are many TCP variants designed for each certain type of networks. As shown in Figure \ref{taxonomy}, \citet{Afanasyev2010} provided an excellent evolutionary graph of most TCP variants based on the problem of which they are trying to solve and how they are behaving. In this paper, high-speed Linux TCP variants that are available for research is presented and explained, as shown in Table \ref{history}, along the following subsections.

\begin{table*}
	\caption {The evolution of High-speed TCP Variants and their implementations in the common operating systems \citep{Afanasyev2010}.}
	\begin{center}
    \begin{tabular}{p{3cm}p{1.8cm}p{3cm}p{2.5cm}p{2cm}l}
    \hline
	TCP Variant	 &	    Year    	&	Base			&	Windows			&	Linux		&	Sun Solaris	\\ \hline
	NewReno		 &	    1999    	&	Reno			&	NA				&	$>$ 2.1.36	&	NA			\\ 
	HS-TCP		 &	    2003    	&	NewReno			&	NA				&	$>$ 2.6.13	&	NA			\\ 
	S-TCP		 &	    2003    	&	NewReno			&	NA				&	$>$ 2.6.13	&	NA			\\ 
	H-TCP		 &	    2004    	&	NewReno			&	NA				&	$>$ 2.6.13	&	NA			\\ 
	BIC-TCP		 &	    2004    	&	HS-TCP			&	NA				&	$>$ 2.6.12	&	NA			\\ 
	TCP-AFRICA	 &	    2005    	&	HS-TCP, Vegas	&	NA				&	NA			&	NA			\\ 
	TCP-Compound &	    2006    	&	HS-TCP, Vegas	&	XP, Vista, Win7	&	$>$ 2.6.14	&	NA			\\ 
	TCP-illinois &	    2006    	&	NewReno, DUAL	&	NA				&	$>$ 2.6.22	&	NA			\\ 
	TCP-FUSION	 &	    2007    	&	Westwood, Vegas	&	NA				&	NA			&	10, 11		\\ 
	YeAH-TCP	 &	    2007    	&	STCP, Vegas		&	NA				&	$>$ 2.6.22	&	NA			\\ 
	TCP-CUBIC	 &	    2008    	&	BIC-TCP			&	NA				&	$>$ 2.6.16	&	NA			\\ \hline    
    \end{tabular}
    \label{history}
    \end{center}
\end{table*}

\subsection{TCP NewReno}
TCP NewReno is a modification of TCP Reno which developed by \citet{Floyd1999} then modified by \citet{Floyd2004, henderson2012} to overcome the problem of Reno's $Fast Recovery$ during the occurrence of multiple packet losses which significantly decreases the Reno's performance in heavy congested networks. In NewReno, the exit from the state of $Fast Recovery$ is only allowed if all the data from the initial congestion window are being acknowledged which senses the $partial$ data ACKs and differentiates it from $new data$ ACKs. More specifically, the $new data$ ACK reception indicates to delivery success of all data which sent before the loss detection while the $partial$ ACK indicates to other losses in the initial congestion window. In fact, NewReno is not designed for high-speed networks \citep{Afanasyev2010}, as shown in Figure \ref{taxonomy}, so it is used here to be compared with the high-speed TCP variants as a benchmark.

\subsection{Scalable TCP (STCP)}
STCP was presented at CERN by \citet{Kelly2003} to overcome the poor performance of the existing congestion control algorithms (such as NewReno) after the increase of bandwidths in high-speed networks. The challenge for this protocol was to achieve better network utilization with higher Bandwidth Delay Products (BDP) without causing any negative impact on the existing traffic. Indeed, STCP is merely sender-side modification to the TCP congestion control algorithm. STCP has been implemented in Linux and then it has provided an improved performance over the gigabit transatlantic network using standard TCP receivers. At that time, the results revealed that, the use of STCP would have a trivial effect on existing network traffic at the same time as enhancing data transfer performance in high-speed networks \citep{Kelly2003}.

The loss-based STCP congestion control algorithm uses $\alpha$, $\beta$ while $(0 < \alpha < 1)$ and $(0 < \beta < 1)$. STCP updates its congestion window after receiving each ACK in a round trip time by $\alpha$, as shown in Equation (\ref{eq1}), in which congestion is not detected but if congestion is detected, it decreases the congestion window by $\beta$, as shown in Equation (\ref{eq2}) \citep{Kelly2003}.  

\begin{equation}
cwnd = cwnd + \alpha
\label{eq1}
\end{equation}

\begin{equation}
cwnd = cwnd - (\beta \ast cwnd)
\label{eq2}
\end{equation}
While $\alpha$ and $\beta$ set to 0.01 and 0.125, respectively.

\subsection{High-speed TCP (HS-TCP)}
\citet{Floyd2003} proposed a new high-speed TCP for large congestion window sizes. This TCP variant was proposed to overcome the poor performance of standard TCP over high-speed networks. HS-TCP is considered as loss-based congestion control algorithm. In fact, HS-TCP did not change the behavior of standard TCP therefore it did not present any risk such as congestion collapse. HS-TCP is merely sender-side modification which increases and decreases congestion window by $\alpha(w)$ and $\beta(w)$, respectively. The resulting functions $\alpha(w)$ and $\beta(w)$ vary from 1 and 0.5, respectively, (when the congestion window is below or equal to 38 packet) to 70 and 0.1, (when the congestion window is greater than 84k packets) \citep{Afanasyev2010,Lar2013}.

Although, HS-TCP succeeded to increase the throughput in high-speed networks, it presented an aggressive behavior than standard TCP which affects its sharing fairness especially when competing with standard TCP flows. Moreover, high-speed TCP presented another problem over high-BDP networks. This problem is known as bursty packet losses which is caused by the standard Slow Start during the phase of an initial Slow Start when an approximate network capacity is not yet determined. In order to overcome this problem, HS-TCP limits its Slow Start to 100 packets. This behavior is well known as ``Limited Slow Start'' which is one of HS-TCP weaknesses.

\subsection{Hamilton TCP (H-TCP)}
H-TCP was presented by \citet{Leith2004} at Hamilton Institute. H-TCP is a loss-based congestion control protocol, which is suitable for high-speed and long distance networks. It is designed to be more fair and effective than conventional TCP. H-TCP defines the increase in the congestion window $w$ as $\alpha(\Delta)$ for each RTT (which increases by a fraction $\alpha(\Delta)/w$ for each reception of non-duplicate ACK), while $\Delta$ is elapsed time since last congestion signal. The final function of the increase is defined as in Equation (\ref{eq3}) \citep{Afanasyev2010,Lar2013}.

\begin{equation}
\alpha(\Delta) = 1 + 10(\Delta - \Delta_{low}) + 0.5 * (\Delta - \Delta_{low})^{2}
\label{eq3}
\end{equation}
Where $\Delta_{low}$ is a predefined value, whenever $\Delta < \Delta_{low}$, $\alpha(\Delta) = 1$. H-TCP reduces its congestion window by $RTT_{ratio}$ Eq.(\ref{eq4}) if $\gamma$ Eq.(\ref{eq5}) is less than 0.2.

\begin{equation}
RTT_{ratio} = \frac{RTT_{min}}{RTT_{max}}
\label{eq4}
\end{equation}

\begin{equation}
\gamma = \left| \frac{B(k) - B(k - 1)}{B(k - 1)} \right|
\label{eq5}
\end{equation}
Where $B(k)$ is the estimation of achieved throughput and $B(k - 1)$ is the estimation of preceding loss event; otherwise it will halve its congestion window.

\subsection{BIC-TCP}
BIC-TCP was presented by \citet{xu2004}, after they had pointed out the problem of RTT-unfairness in HS-TCP and STCP. More specifically, assume that two TCP flows are sharing one bottleneck and they detect the loss synchronously, if the two flows are HS-TCP, the flow that its RTT is $x$ times smaller can have a network share of $x^{4.56}$ times larger. But if two STCP flows are used, the smaller RTT will grab all the \mbox{network} \mbox{bandwidth} while the higher RTT will get nothing. Hence, BIC-TCP was presented to solve this problem of absolute RTT-unfairness \citep{Harfoush2004, Afanasyev2010}.

Despite of the improved performance of BIC-TCP, its function of window growth can be highly aggressive especially over low-speed or short-distance networks. Furthermore, BIC-TCP may achieve a bad inter-fairness and RTT-fairness due to its dependability on RTT measurements. As well as, it has a high complexity due to the several modes (binary search increase, max probing, Smax and Smin) of the algorithm itself. Thus, \mbox{BIC-TCP} has been reviewed and modified in CUBIC which conserves the stability and scalability of \mbox{BIC-TCP}, decreases the complexity, and increases the fairness \citep{Ha2008,Afanasyev2010}.

\subsection{TCP Africa}
TCP-Africa (Adaptive and Fair Rapid Increase Congestion Avoidance) was presented by \citet{King2005}. TCP-Africa was designed to solve the problems that were appearing in high-BDP networks. The aggressiveness and scalability of HS-TCP (in case of congestion-free) and the conservative attribute of standard NewReno (in case of congestion) have been combined to gain a better performance than the existing TCP variants. TCP-Africa is loss-delay-based algorithm, which has borrowed its behavior $(congestion/non-congestion)$ from TCP Vegas algorithm; by comparing the estimated buffer of the network $\Delta$ to a predefined constant $\alpha$. In TCP-Africa, when $(\Delta < \alpha)$ which indicates to a little buffering space, it switches to $fast mode$ and immediately applies the Congestion Avoidance and Fast Recovery of HS-TCP algorithm. In this case, the decrease and increase steps are calculated as $\beta(w)$ and $\alpha(w)$, respectively. Otherwise, it switches to $slow mode$ which applies the rules of NewReno that increases by $one$ after every $ACK$ reception and decreases by $halving$ the congestion window after loss detection. 

TCP-Africa has been evaluated by simulation and presented good bandwidth utilization in high-BDP networks \citep{King2005,Afanasyev2010}. It showed a lower loss ratio than HS-TCP and STC. It also presented high fairness (RTT-, intra-, inter-) similar to that presented by NewReno. In despite of that improvement, TCP-Africa has not been implemented in real operating systems, whereas a similar $multiple-mode$ congestion algorithm which is Compound TCP has been implemented in Microsoft Windows operating systems \citep{Afanasyev2010,Lar2013}.

\subsection{TCP-illinois}
TCP-illinois was introduced at UIUC by \citet{Liu2008}. It is a sender-side protocol which modifies $AIMD$ algorithm of the standard TCP (Reno, NewReno or Sack). It uses $loss$ and $delay$ as congestion signals to increase or decrease its congestion window. TCP-illinois achieves better performance than the standard TCP and shares the network bandwidth fairly especially over high-BDP networks. TCP-illinois updates its congestion window after every $ACK$ reception in a round trip time by $(\alpha / cwnd)$ in which congestion has not detected but when congestion detected, TCP-illinois decreases its congestion window by $(\beta * cwnd)$ as in Equations (\ref{eq8}) and (\ref{eq9}) \citep{Liu2008}, respectively.

\begin{equation}
cwnd = cwnd + (\alpha / cwnd)
\label{eq8}
\end{equation}

\begin{equation}
cwnd = cwnd - (\beta * cwnd)
\label{eq9}
\end{equation}

TCP-illinois uses loss signal to set the direction and use delay to calculate the step of window size change by $f_1(.)$ and  $f_2(.)$ as explained in reference \citep{Liu2008}, while $(0 \leq \alpha \leq 1)$, $(0.125 \leq \beta \leq 0.5)$ and $\alpha = f_1(d_a)$, $\beta = f_2(d_a)$, where $(d_a)$ is delay-average.

\subsection{Compound TCP (C-TCP)}
\citet{Tan2006} introduced new loss-delay-based TCP variant named C-TCP. As TCP-Africa, C-TCP combines two modes of NewReno and HS-TCP to increase the bandwidth utilization over high-BDP networks. C-TCP compares $\alpha$ to the estimated $\Delta$, where $\alpha$ is small predefined constant. When $\Delta$ exceeds $\alpha$, C-TCP gently reduces $W_{fast}$ by a predefined $\zeta$ as shown in Equation (\ref{eq11}) \citep{Afanasyev2010}. 

\begin{equation}
W_{fast} = W_{fast} - (\zeta * \Delta)
\label{eq11}
\end{equation}

C-TCP calculates $W_{fast}$ to add it to the final congestion window as shown in Equation (\ref{eq10}) \citep{Afanasyev2010}.

\begin{equation}
W = W_{reno} + W_{fast}	
\label{eq10}
\end{equation}

This $W_{fast}$ is a smooth movement from HS-TCP $fast mode$ to NewReno $slow mode$. C-TCP behavior is very similar to TCP-Africa but C-TCP shows a convex curve after exceeding the threshold while TCP- Africa shows linear increase. In despite of the changes in its behavior, C-TCP still achieves as same performance as TCP-Africa and even presents another problem of RTT estimation which is inherited from TCP Vegas. This problem makes C-TCP very sensitive to RTT measurements which makes it slightly unfair. However, C-TCP is currently the most deployed congestion control algorithm since its implementation in Microsoft Windows operating systems \citep{Afanasyev2010}. 

\subsection{YeAH TCP}
YeAH (Yet Another High-speed) TCP was presented by \citet{Baiocchi2007}. It is similar in spirit of TCP-Africa and C-TCP. It combines loss detection and RTT estimation to predict network delay. Similarly, YeAH combines NewReno and STCP instead of HS-TCP, so it increases the congestion window by $one$ every RTT and $halving$ it if a loss is detected (by receiving three duplicated ACKs). More specifically, if $(\Delta < \alpha)$, where $\alpha$ is a predefined threshold, and $(Q/RTT_{min} < \phi)$, where $\phi$ is another predefined threshold, YeAH switches to $fast mode$ and behaves similarly as STCP. Otherwise, a $slow mode$ of NewReno is applied. Briefly, YeAH showed higher efficiency and fairness (inter-, intra-, RTT-) than TCP-Africa and C-TCP especially in high-BDP networks but it still has the same problem of RTT estimation which is inherited from Vegas \citep{Afanasyev2010,Lar2013}. 

\subsection{TCP Fusion}
\citet{Kaneko2007} presented TCP Fusion which combining $Westwood's$ $achievable$ $rate$, $DUAL’s$ $queuing$ $delay$, and $Vegas’$ $used$ $network$ $buffering$ $estimations$. Depending on the absolute threshold value of queuing delay, Fusion switches to its three modes; if the queuing delay is lower than the predefined threshold, the $fast mode$ is applied which increases its $cwnd$ by a predefined achievable rate estimation fraction of Westwood. While if the current queuing delay is greater than $three$ times of the threshold, $cwnd$ is decreased by the number of buffered packets in the network. Otherwise, if the queuing delay is somewhere between $one$ and $three$ times of the predefined threshold, Fusion keeps its $cwnd$ as it is. Indeed, experimental results showed the improvement of Fusion performance metrics such as bandwidth utilization and fairness compared to C-TCP, HS-TCP, BIC and Fast. Despite of the improvement, Fusion has many limitations such as the problem of predefining the threshold which is done manually, and the more critical problem which may lead Fusion in some cases to behave similar to standard NewReno most of the time \citep{Afanasyev2010}. 

\subsection{CUBIC TCP}
CUBIC TCP was presented by \citet{Ha2008} ant it is the current default TCP algorithm in most Linux operating systems. It modified the $linear$ function of $cwnd$ increase in the existing TCP variants to $cubic$ function in order to enhance its scalability over high-BDP networks. \citet{Ha2008} have reviewed BIC algorithm to come up with CUBIC which borrowed the $cubic$ function of congestion window from H-TCP approach as shown in Equation (\ref{eq12}) \citep{Afanasyev2010}.

\begin{equation}
w = C \left(\Delta - \sqrt[3]{\frac{\beta * w_{max}}{C}}\right)^3 + w_{max}
\label{eq12}
\end{equation}
where $C$ is a predefined constant, $\beta$ is a coefficient of multiplicative decrease in Fast Recovery, and $w_{max}$ is the congestion window size just before the last registered loss detection. $Limited$ $Slow$ $Start$, $Rapid$ $Convergence$ and $RTT$ $independence$ in CUBIC, all provided higher fairness (RTT-, intra-) and higher scalability. The target window $w_{max}$ is calculated in the initial stage of the window increase which is discovered by the $right$ branch of $cubic$ function. The exponential increase of standard Slow Start is more aggressive than the discovery of the window increase which is more scalable in high-BDP networks. Upon loss detection, if this loss is temporary and $w_{max}$ is not reached yet, $cwnd$ will be increased according to both $right$ and $left$ branches of the $cubic$ function. 

Moreover, CUBIC ensures that, its throughput is not lower than the throughput of the standard NewReno, which is done  by enforcing the calculated value of $w_{reno}$ whenever $w_{max}$ is going lower than $w_{reno}$. This complicated behavior of CUBIC algorithm confirms a very high performance and fairness attributes, which make it the second most used TCP variant after being the standard TCP of Linux operating systems. However, CUBIC is still have some limitations that lead to under-utilization of the available bandwidth and produces a huge number of packet losses especially in high-BDP networks. These limitations are due to the dependency of $loss$ which is the only congestion signal used in this algorithm \citep{Afanasyev2010,Lar2013,Ha2008}.

\subsection{Latest Issues}

\citet{fu2007}, \citet{alrshah2009}, \citet{qureshi2012} and \citet{alrshah2013} confirmed that, the single-based TCP with an appropriate modification can overcome and well replace the parallel-based TCP and it may be able to fully utilize the high-speed bandwidths. While \citet{ha2011} mentioned that, standard Slow Start becomes inappropriate for the high-BDP networks and they stated two reasons for this problem as below: 
\begin{enumerate}
\item The exponential increase of the congestion window results a heavy packet losses that make the entire system completely unresponsive for a long period of time during the loss recovery stage.
\item Some optimizations, that applied to Slow Start, happen to slow down the loss recovery followed by Slow Start which leads to under-utilization of  the network resources.
\end{enumerate}

In order to solve the above mentioned problems, they presented a new Slow Start algorithm named "HyStart". This algorithm finds a safe exit point from Slow Start to Congestion Avoidance without causing a heavy packet losses. This algorithm improves the throughput of TCP and it has been already applied to CUBIC since Linux 2.6.29 as a default Slow Start. \citet{xu2011} proposed a new hybrid congestion control called HCC-TCP which is loss-delay-based. HCC-TCP improves the throughput and fairness as well.

\citet{dangi2012a} proposed a new hybrid (loss-delay-based) congestion control scheme. The experiments of \citet{dangi2012b} reveals that, HCC-TCP can achieve an efficient performance on throughput over high-BDP networks. In order to increase the bandwidth utilization, \citet{khalil2012} proposed a new congestion control scheme called $Swift-Start$. It changes the way of estimating the available bandwidth to avoid the congestion which caused by over or under bandwidth estimation. \citet{cavendish2012} prove that TCP can achieve a superior performance if its parameters are tuned well depending on network and path conditions.

Moreover, network buffers are going towards the $near-zero$ buffer, as mentioned in \citep{enachescu2006buffer1, beheshti2006buffer2, prasad2007buffer3, vishwanath2008buffer5, vishwanath2009buffer6, vishwanath2009buffer4, vishwanath2009buffer8, sivaraman2009buffer9, legrange2009buffer10, vishwanath2011buffer7}, to fit the all-fiber networks which is the fastest type of high-speed networks yet. Consequently, it is very important to take the case of $near-zero$ buffer network into account in the future TCP performance evaluation.

As mentioned above, TCP is still suffering from many problems, and researchers are still modifying and improving it. Some researchers mix different modes, such as fast and slow modes, and switch between them based on the state of the network. And some researchers mix different approaches, such as loss and delay based approaches, to improve the performance. And some researchers are estimating the RTT and bandwidth to avoid the severe congestion which can lead to congestion collapse. While some are trying to modify the algorithm itself by modifying Slow Start or Congestion Avoidance algorithm, and some of the rest are trying to tune the TCP parameters carefully to achieve a superior performance.

\section{Performance Evaluation of TCP Variants}
\label{PE}
In this paper, two simulation-based experiments have been conducted to show the performance differences among high-speed Linux TCP variants over high-BDP congested and non-congested networks. The first experiment has been conducted to evaluate the performance of TCP over non-congested network to mimic the ideal case of the network, then, to show the ability of TCP on bandwidth utilization, and to determine the points of weaknesses in its mechanism. In addition to that, the second experiment has been conducted to evaluate the performance of TCP over congested bottleneck in order to simulate a real network scenario.

In the first experiment TCP variants have been evaluated by measuring the average throughput and loss ratio while in the second experiment they have been evaluated by measuring the average throughput, loss ratio, intra-fairness and RTT-fairness. More specifically, measuring the average throughput is beneficial to show the ability of link utilization, while measuring the loss ratio is helpful to show the quantity of lost data which negatively affects the general performance of TCP. On the other hand, measuring (intra-, RTT-) fairness is to show the quality of sharing the link between the competing TCP flows based on Jain's fairness index \citep{jain1984}. All of these measurements are conducted to show the advantages and disadvantages of all involved TCP variants to determine the points of strengths and weaknesses of every TCP variant in order to help the process of improving the performance of these variants.

\subsection{Experiments Setup}

\begin{figure}[hbtp]
\includegraphics[width=\linewidth]{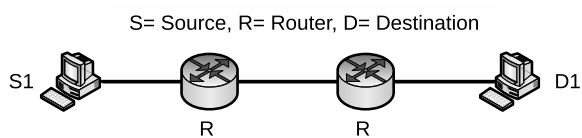}
\caption{Non-congested network topology.}
\label{topology-ideal}
\end{figure}

In the first experiment, a standard single dumbbell topology has been used as shown in Figure \ref{topology-ideal}. Only one sender ($S1$) and one receiver ($D1$) are used. $S1$ sends data to $D1$ through two routers on the path. $S1$ and $D1$ are connected to the routers over LAN with $1Gbps$ speed and $1ms$ propagation delay. While the routers are linked by $1Gbps$ speed with a propagation delay of $100ms$. As it is clear in this topology, this network does not have bottleneck, therefore it is considered as the best case of TCP over an ideal network.

\begin{figure}[hbtp]
\includegraphics[width=\linewidth]{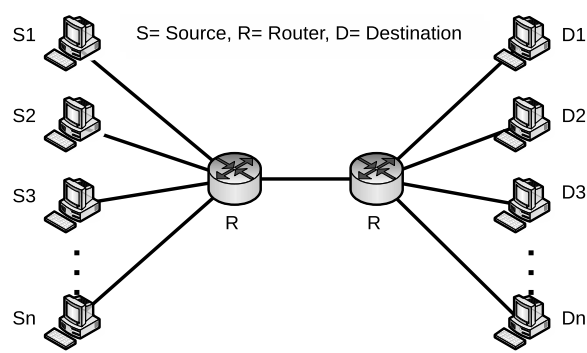}
\caption{Network topology with standard dumbbell bottleneck.}
\label{topology}
\end{figure}

As for  the second experiment setup, a standard single dumbbell topology has been used as shown in Figure \ref{topology}.  As shown in the network topology, there are $n$ competing senders ($S1$, $S2$, $S3$, ..., $Sn$) send data synchronously to $n$ receivers ($D1$, $D2$, $D3$, ..., $Dn$) through a shared single bottleneck. All nodes of sources and destinations are connected to bottleneck routers over LAN with $1Gbps$ speed and $1ms$ propagation delay. While the bottleneck link is $1Gbps$ speed with a propagation delay of $100ms$. Consequently, the proper bandwidth of the shared bottleneck, which is needed by the concurrent senders, is $4Gbps$ while the available is only $1Gbps$, this in order to simulate a real congested bottleneck. 

This experiment is repeated for every TCP variant separately with different buffer sizes which starts from $100$ to $5000$ packets. In fact, this experiments show the impact of bottleneck congestion and buffer size on the performance of the examined TCP variants and also show the performance changes when a smaller buffer size is applied. Scalable, HS-TCP, BIC, H-TCP, CUBIC, Africa, Compound, Fusion, NewReno, illinois and YeAH are involved in these experiments. All of these TCP variants are added into NS2 version 2.35 which is installed on Linux openSuse 12.2, kernel version 3.4.28 over Intel Core-i7 machine to perform this simulation-based comparison. Table \ref{params} shows the experiment setup and the simulation parameters.

\begin{table}[t!]
	\caption{Experiment Parameters.}
	\begin{center}
	\begin{tabular}{p{0.15cm}p{2.24cm}p{4.16cm}} 
	\hline
	No.& Parameter				 				&	Value														\\ \hline
	1. & TCP Variants		 	 				&	Scalable, HS-TCP, BIC, H-TCP, CUBIC, Africa, C-TCP, Fusion,  NewReno, illinois and YeAH.\\ 
	2. & Link capacity							&	1000 Mbps for all.											\\ 
	3. & Link delay					 			&	1ms node to router.											\\ 
	   & 									 	&	100ms router to router.										\\ 
	4. & BDP					 				&	12750KB (High-BDP as in \cite{RFC1072}).					\\ 
	5. & Buffer size			 				&	from 100 to 5000 packets.									\\ 
	6. & Packet size			 				&	1000 bytes.													\\ 
	7. & Queuing Algo	 						&	Drop Tail.													\\ 
	8. & Traffic type			 				&	FTP															\\ 
	9. & Simulation time		 				&	100 seconds.												\\ \hline
	\end{tabular}
	\label{params}
	\end{center}
\end{table}

\subsection{Results and Discussion}
Based on the results of the first experiment, it is briefly concluded that, TCP Slow Start has a fatal problem known as burst loss. Burst loss happens when  TCP jumps exponentially to reach the maximum $cwnd$ in order to quickly utilize the bandwidth of the network. The lack of the information about the link bandwidth makes TCP increases its $cwnd$ until it detects the first loss then it halves its $cwnd$ and enters the linear increase stage. In other words, the burst loss happens when the first loss is detected. Indeed, this burst loss can severely affect the performance of TCP and it may even lead to congestion collapse.

\begin{figure*} [t!]
	\centering
	\begin{center}
		\subfigure[TCP NewReno] 
		{
			\includegraphics[scale=0.413]{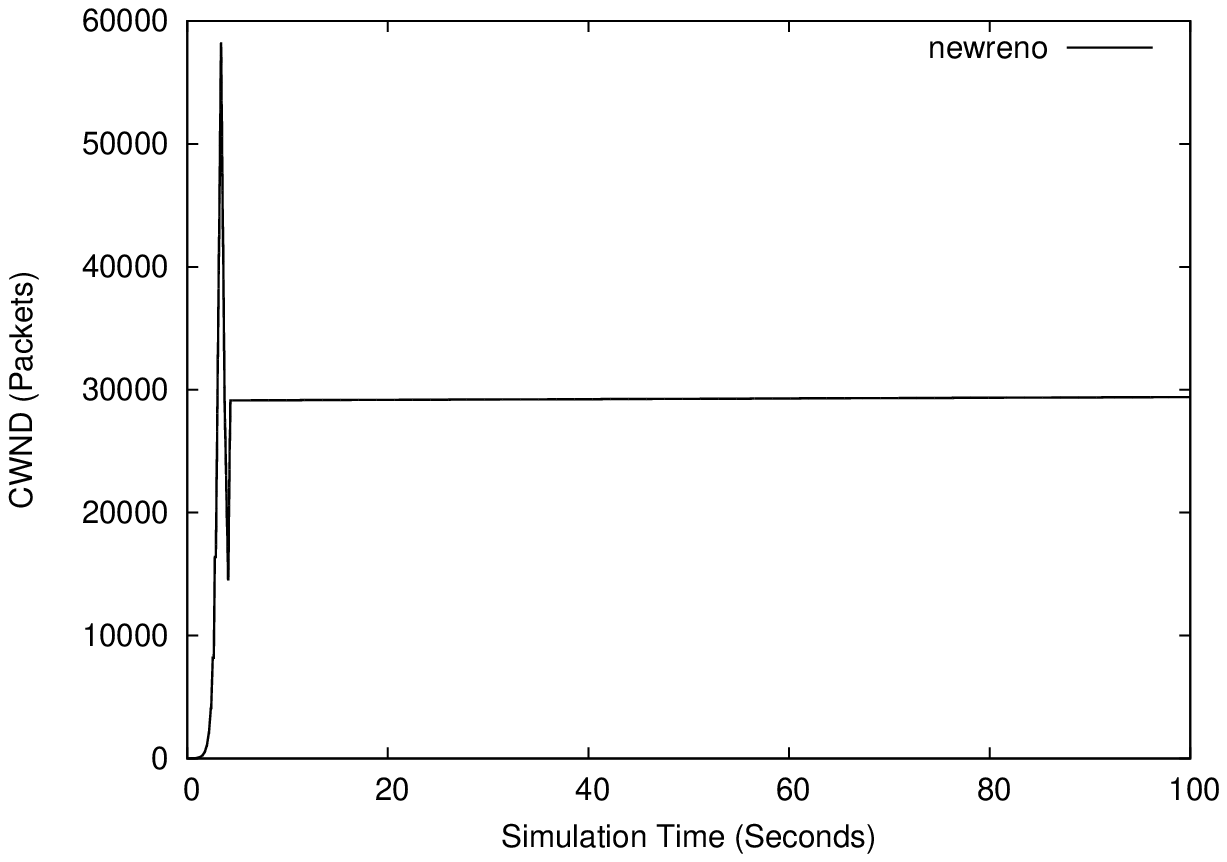}
			\label{NewReno}
		}
		\subfigure[TCP Africa] 
		{
			\includegraphics[scale=0.413]{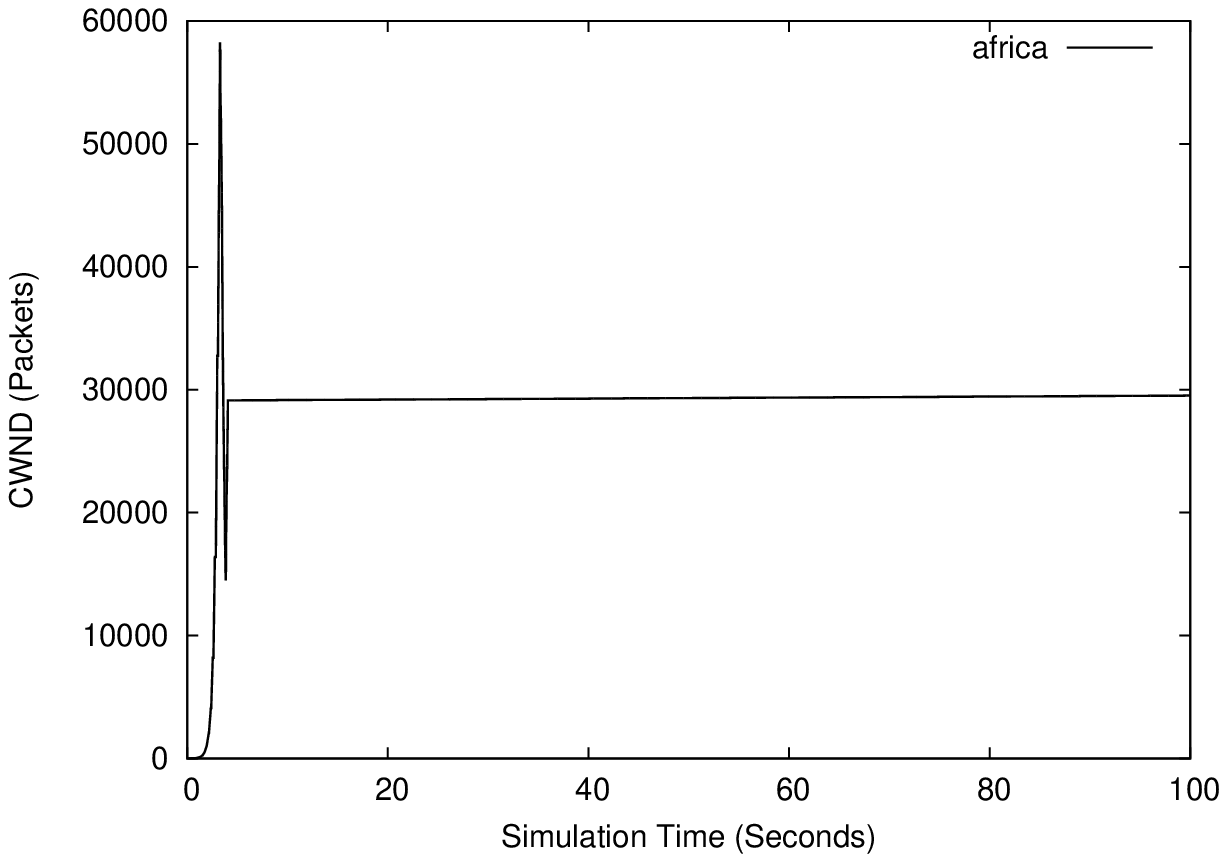}
			\label{africa}
		}		
		\subfigure[TCP illinois] 
		{
			\includegraphics[scale=0.413]{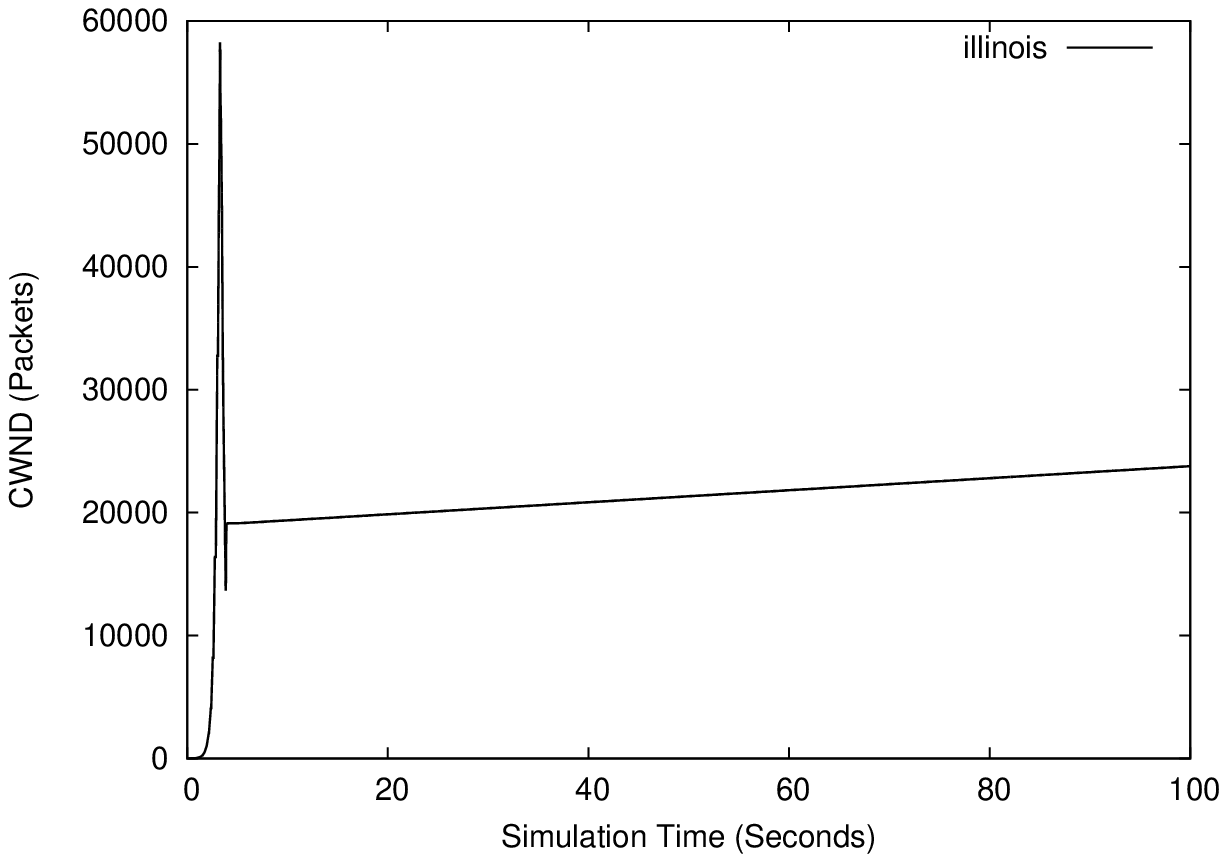}
			\label{illinois}
		}		
		\subfigure[Compound TCP] 
		{
			\includegraphics[scale=0.413]{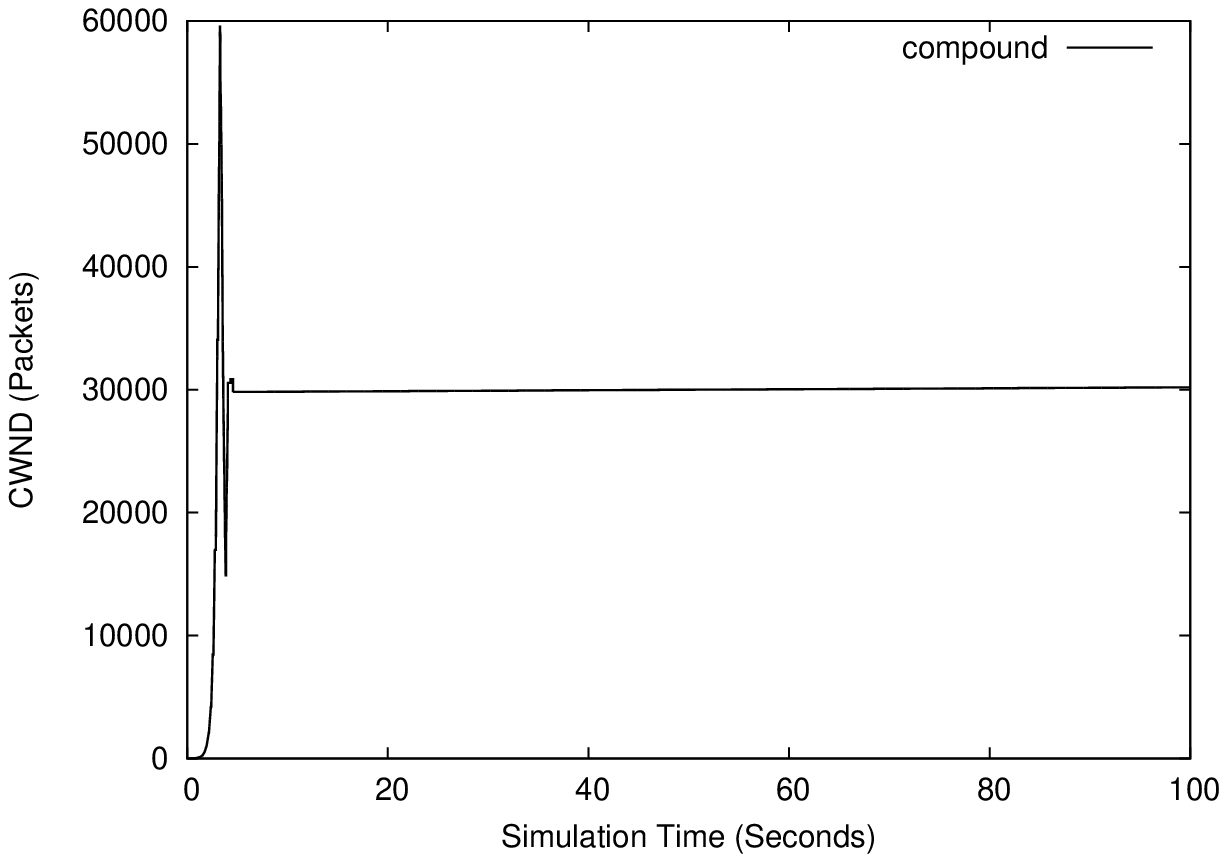}
			\label{compound}
		}		
		\subfigure[TCP Fusion] 
		{
			\includegraphics[scale=0.413]{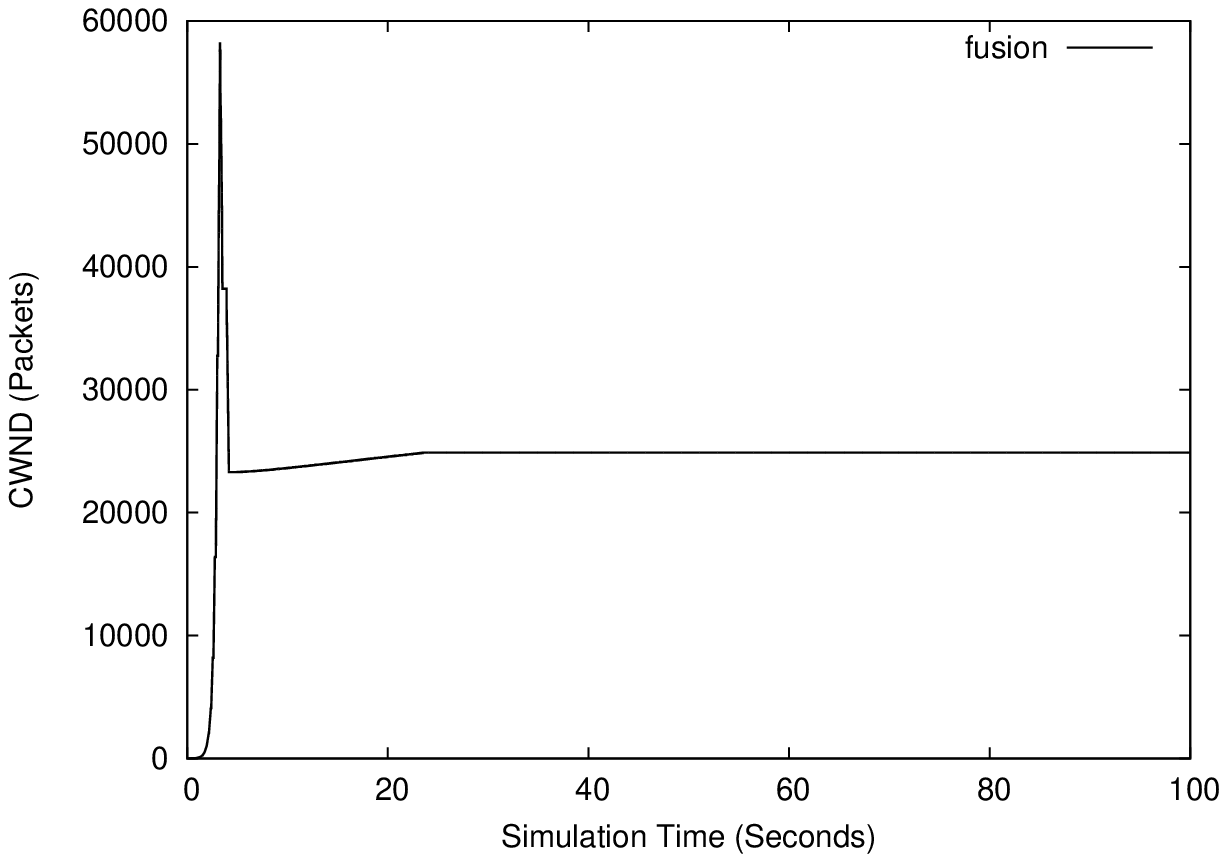}
			\label{fusion}
		}		
		\subfigure[Scalable TCP] 
		{
			\includegraphics[scale=0.413]{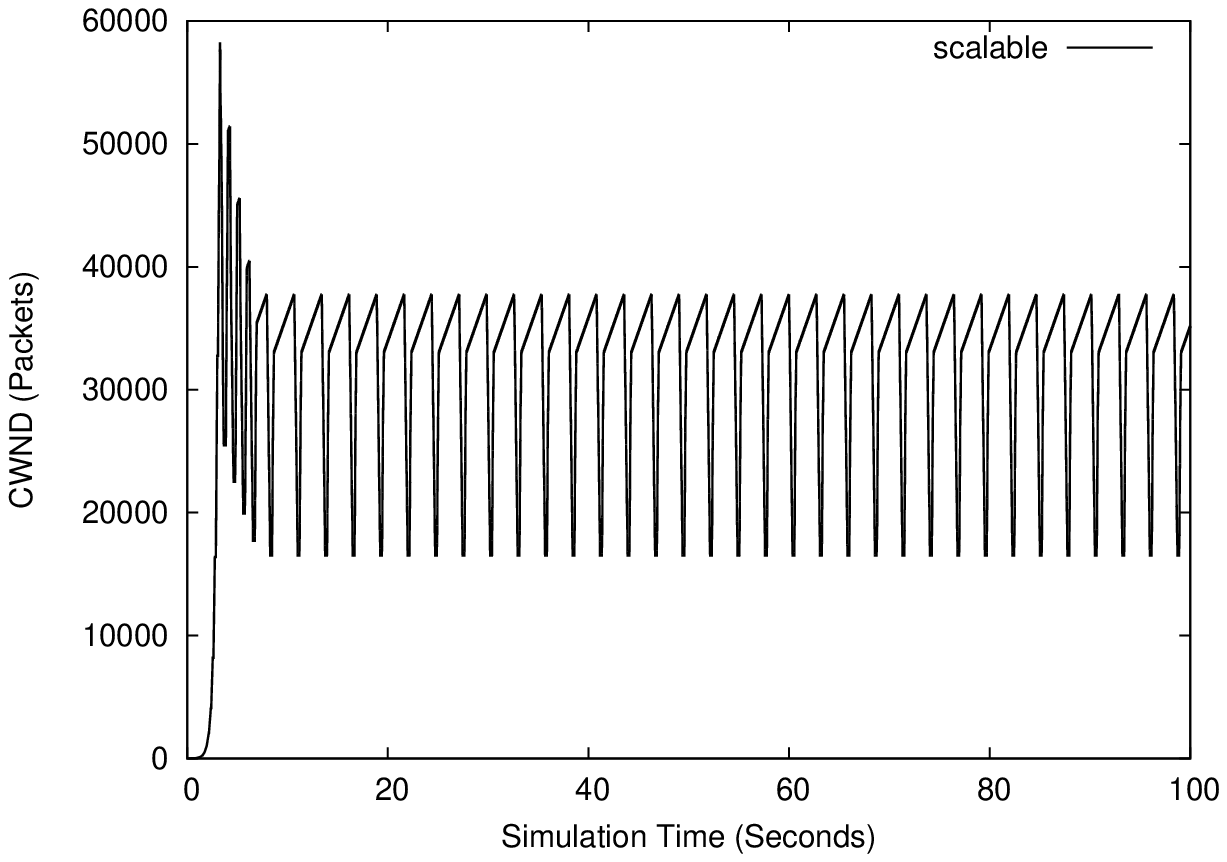}
			\label{stcp}
		}		
		\subfigure[High-speed TCP] 
		{
			\includegraphics[scale=0.413]{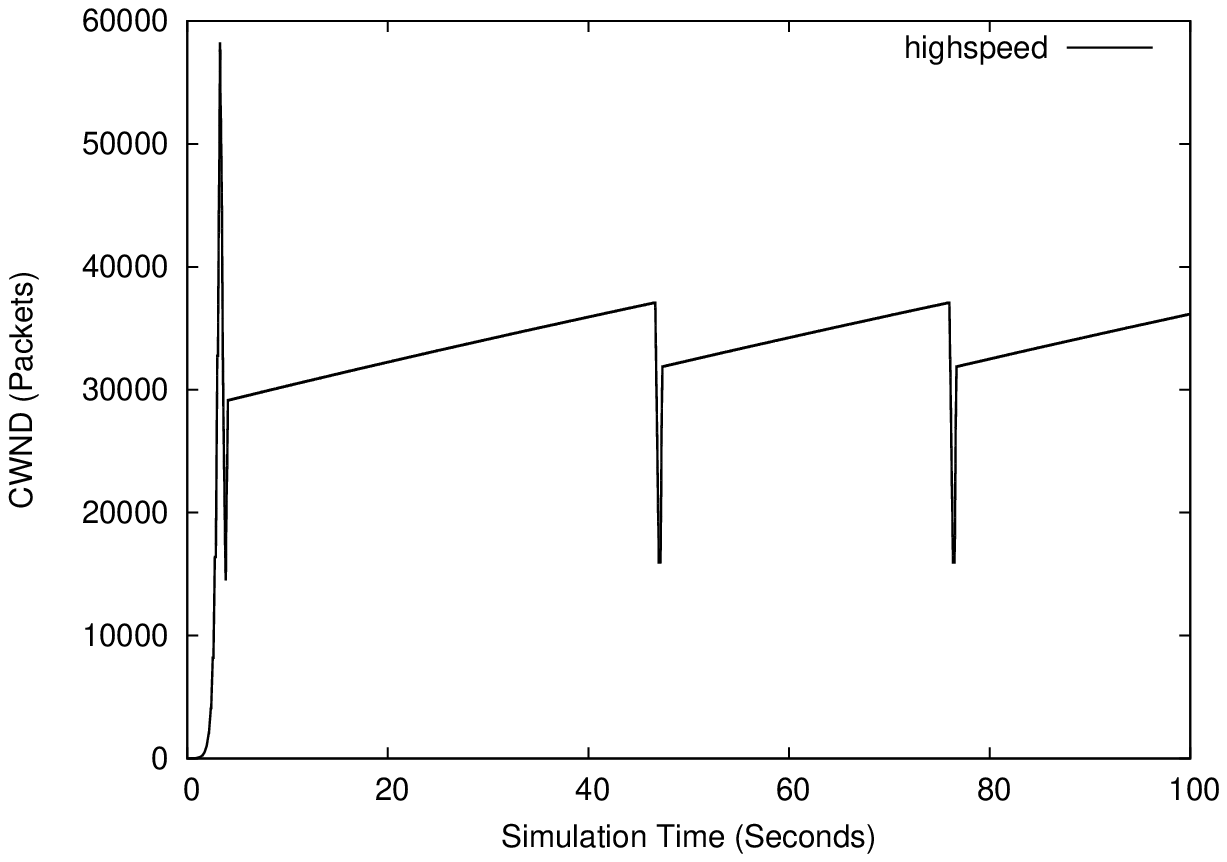}
			\label{hstcp}
		}
		\subfigure[Hamilton TCP] 
		{
			\includegraphics[scale=0.413]{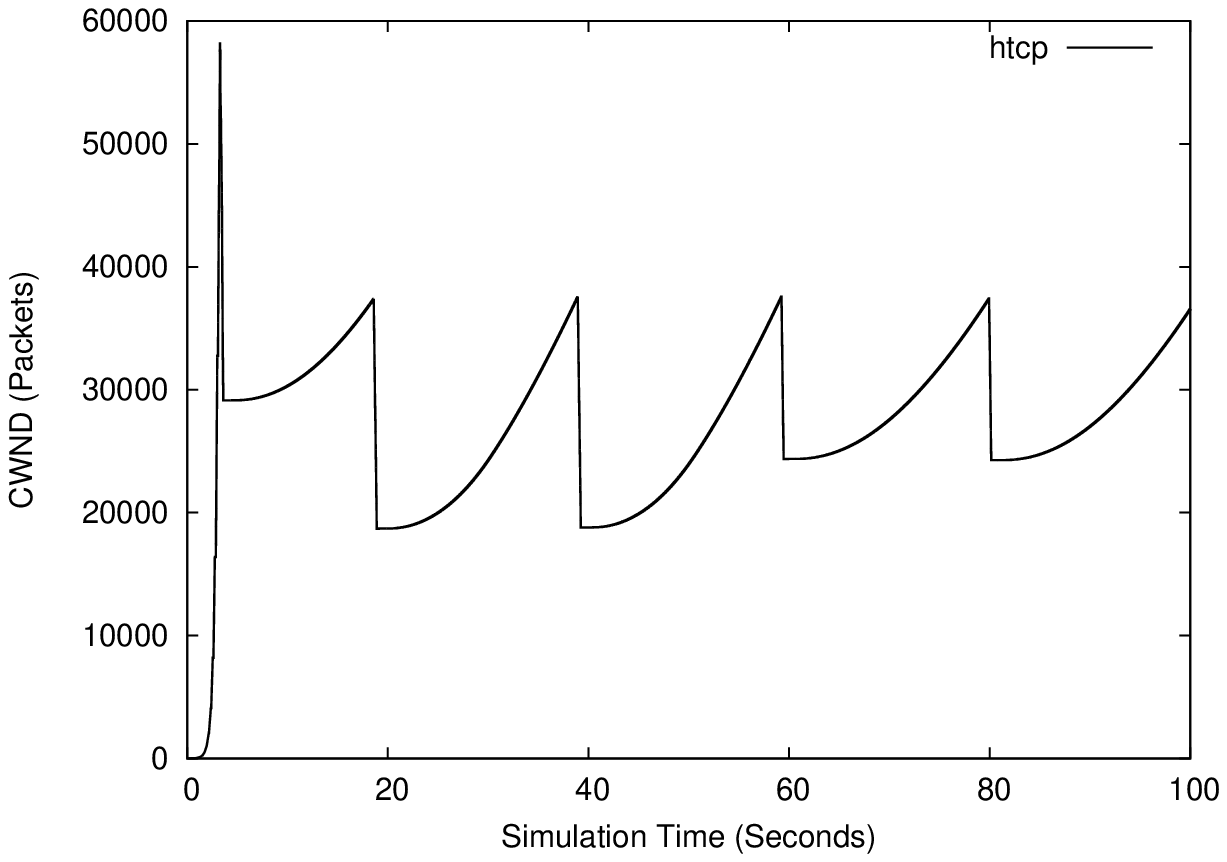}
			\label{htcp}
		}		
		\subfigure[YeAH TCP] 
		{
			\includegraphics[scale=0.413]{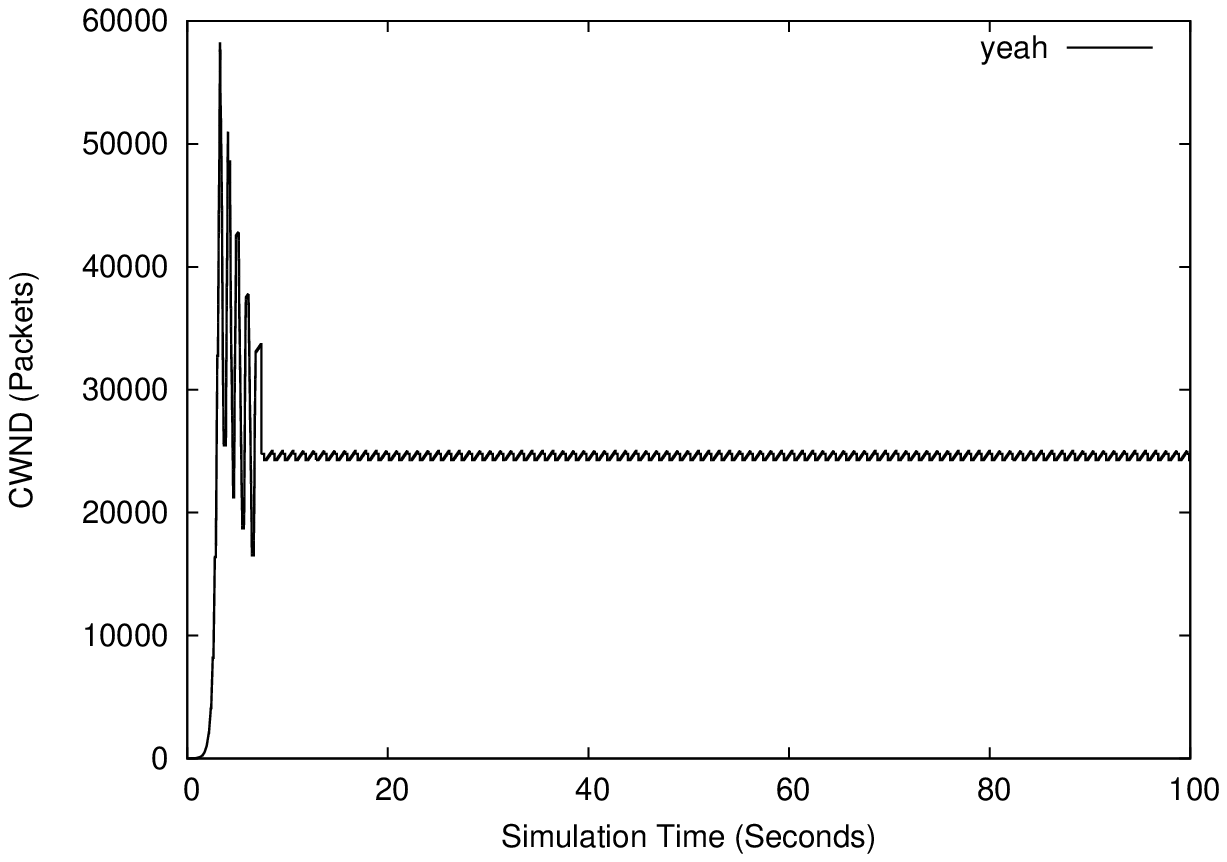}
			\label{yeah}
		}		
		\subfigure[Bic TCP] 
		{
			\includegraphics[scale=0.413]{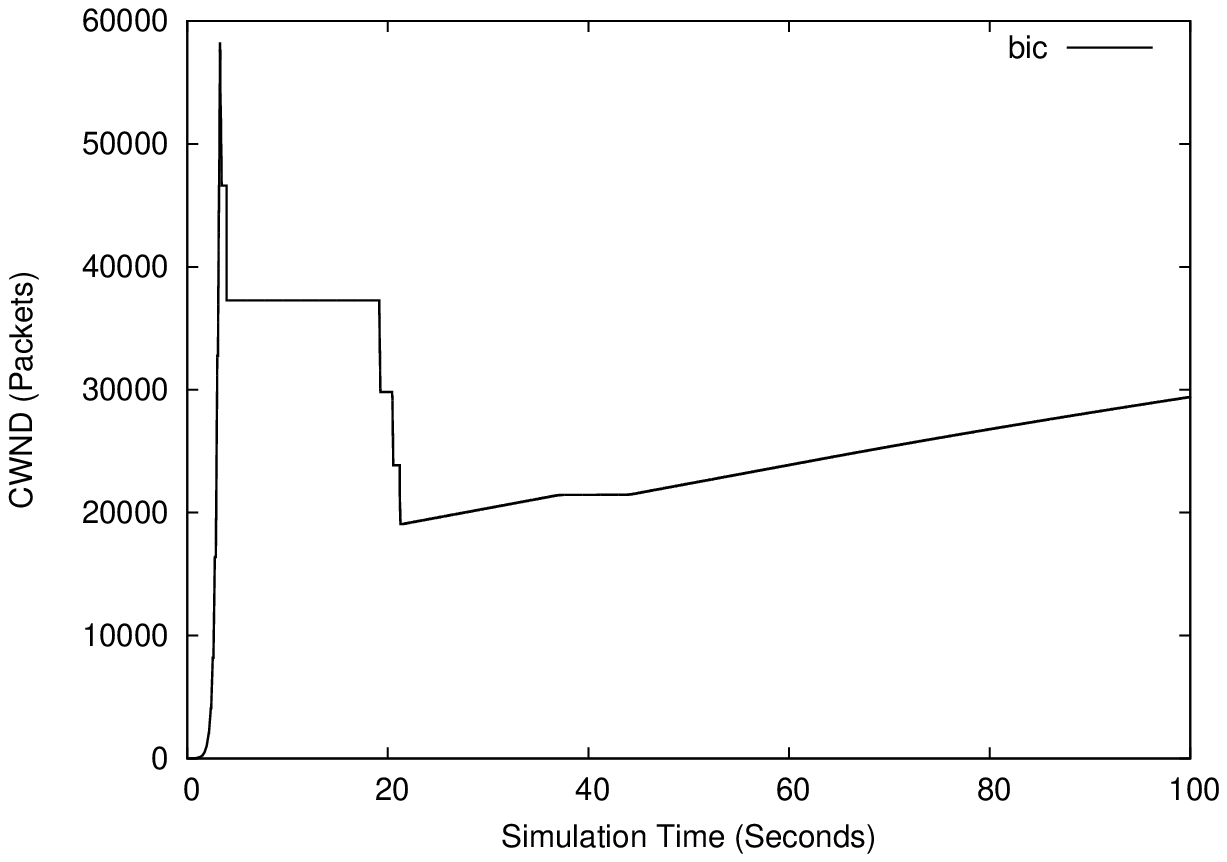}
			\label{bic}
		}
		\subfigure[Cubic TCP] 
		{
			\includegraphics[scale=0.413]{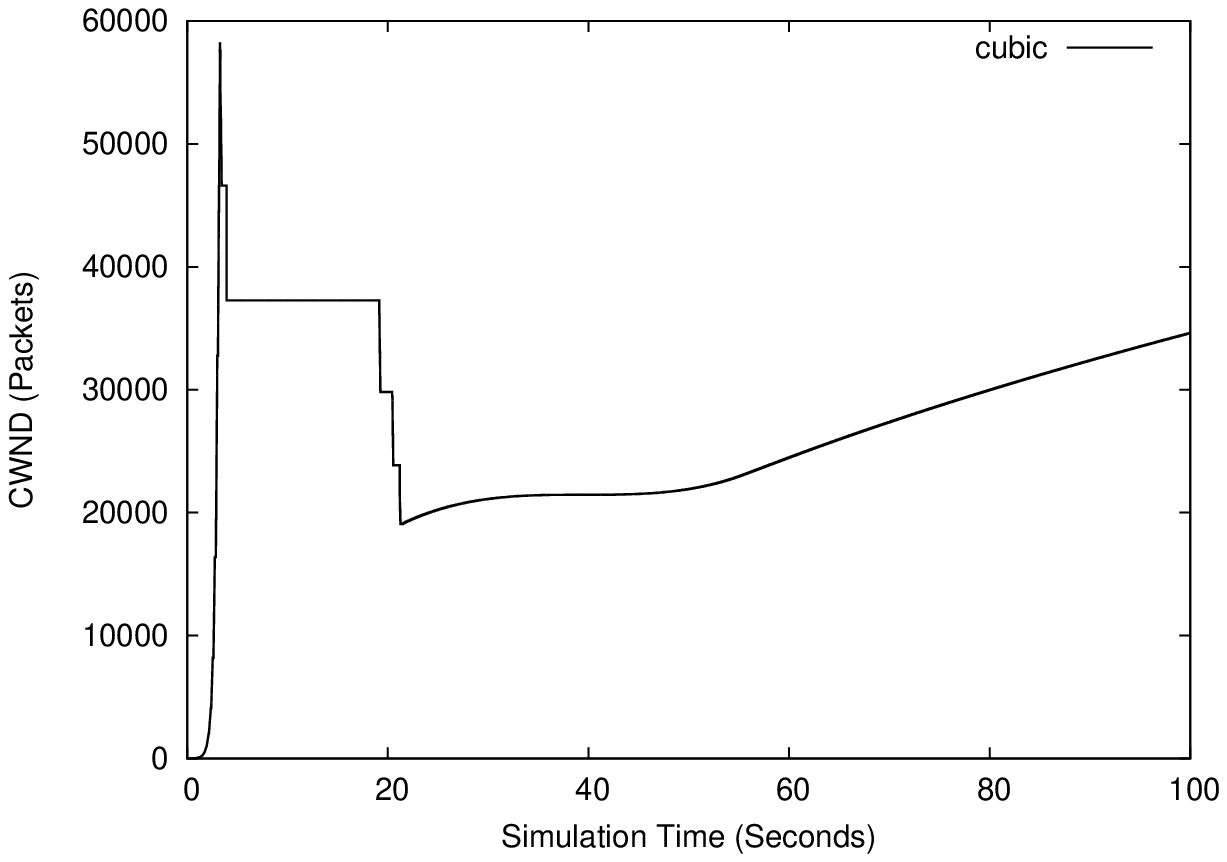}
			\label{cubic}
		}		
	\end{center}
	\caption{The congestion window dynamics of the examined TCP variants.}
	\label{fig:cwnd-dynamics}
\end{figure*}

All the TCP variants in this experiment, jump to reach a $cwnd$ of around 60000 packets but after the first loss is detected, their behaviors become completely different. NewReno, Africa, illinois, C-TCP and Fusion drop their $cwnd$ to the half then increase it linearly in a very slow manner as shown in Figures \ref{NewReno}, \ref{africa}, \ref{illinois}, \ref{compound} and \ref{fusion}, respectively. This linear increase behavior consumes a long time to reach the upper limit of the network bandwidth again which results an under-utilization of the network resources.

Diversely, STCP, HS-TCP, H-TCP and YeAH drop their $cwnd$ to the half then increase it in an oscillating manner as shown in Figures \ref{stcp}, \ref{hstcp}, \ref{htcp}, \ref{yeah}, respectively. This behavior increases the bandwidth utilization to some extent, but some of these TCP variants such as STCP, HS-TCP and H-TCP are still suffer from the problem of under-utilization while YeAH has achieved higher bandwidth utilization due to its conservative reduction in the Congestion Avoidance stage. More specifically, all of the aforementioned TCP variants reduce their $cwnd$ to the half when the loss is detected, but in YeAH, $cwnd$ is reduced to the half if the loss is detected in Slow Start stage while it reduced gently in the Congestion Avoidance stage based on the changes of network delay.

More differently, BIC and CUBIC drop their $cwnd$, then increase it in a rapid convergence manner as shown in Figures \ref{bic} and \ref{cubic}, respectively. Indeed, they reduce their $cwnd$ to around 85\% of the last $cwnd$ whenever a loss is detected regardless of the loss is coming from Slow Start or Congestion Avoidance. This behavior results in a higher bandwidth utilization than the other TCP variants which halving their $cwnd$ after every loss. As a final point, and based on observation, the conservative reduction of the $cwnd$ can help TCP to increase: (1) the bandwidth utilization by releasing a small part of the used bandwidth, after every loss, which helps to reach again to the higher limit more faster. (2) the fairness by reducing the gab between the max limit of the bandwidth and the reduced $cwnd$ after loss detection.

\begin{figure}[t!]
\includegraphics[scale=0.6]{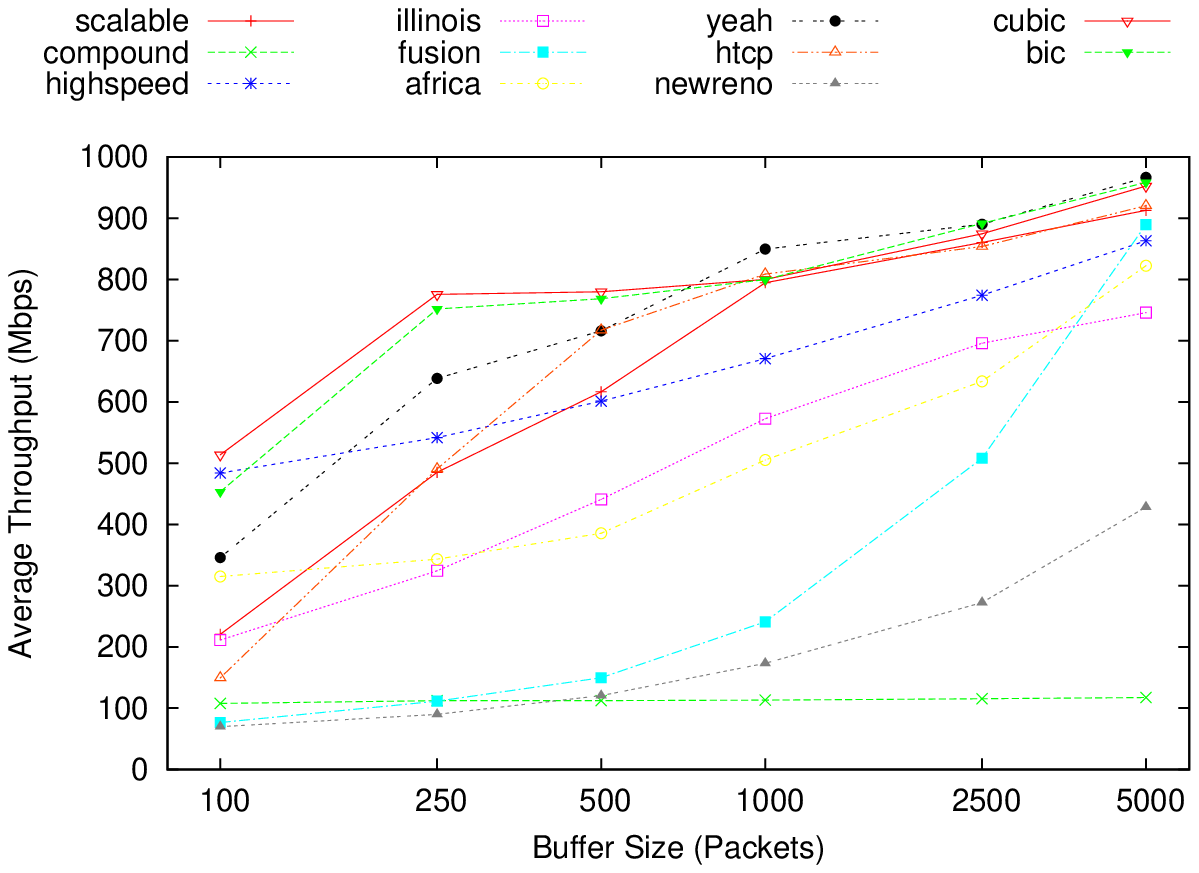}
\caption{Average Throughput vs. Buffer Size}
\label{avrg-thrghpt}
\end{figure}

As for the results of the second experiment, Figure \ref{avrg-thrghpt} shows that, CUBIC, BIC and YeAH are the best three TCP variants in terms of throughput. More specifically, in the cases of 5000, 2500 and 1000 packets buffer size YeAH is a bit better than CUBIC and BIC. While in the other cases of smaller buffer sizes, CUBIC overcomes all other TCP variants whenever buffer size is going closer to $near-zero$ buffer. As mentioned above, YeAH, BIC and CUBIC are using a conservative and gentle reduction of $cwnd$ which helps them to outperform the other variants of TCP that half their $cwnd$ whenever a loss is detected.

\begin{figure}[t!]
\includegraphics[scale=0.6]{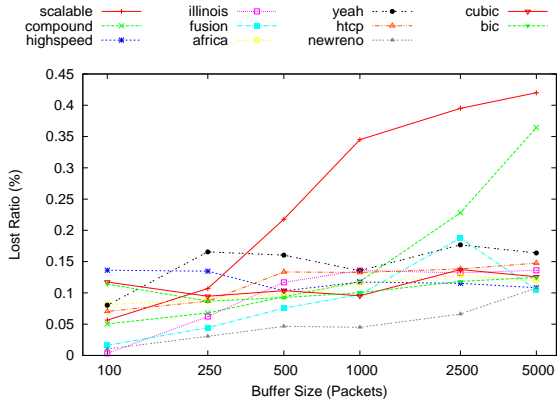}
\caption{Loss Ratio vs. Buffer Size}
\label{lostratio}
\end{figure}

Indeed, average throughput solely is not enough to evaluate the performance of TCP variants. The other important performance metric, which is necessary to evaluate the performance of TCP, is the loss ratio. In Figure \ref{lostratio}, the lowest loss ratio is NewReno, which has a very poor throughput average because it is not designed for high-speed networks. For this reason it will not be considered. Unlikely, scalable and compound have the highest loss ratio. While, BIC, CUBIC and H-TCP are keeping their loss ratio stable regardless of the changes in buffer size. In fact, all of the compared TCP variants produce similar number of lost packets, but when the count of lost packets compared to the total sent data as a loss ratio, it will give a completely different readouts.

\begin{figure}[t!]
\includegraphics[scale=0.6]{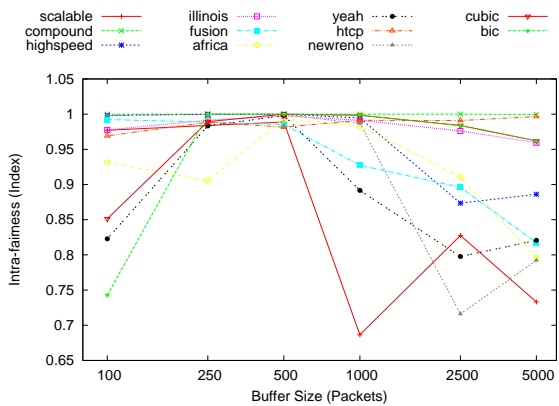}
\caption{Intra-fairness vs. Buffer Size}
\label{Intrafair}
\end{figure}

\begin{figure}[t!]
\includegraphics[scale=0.6]{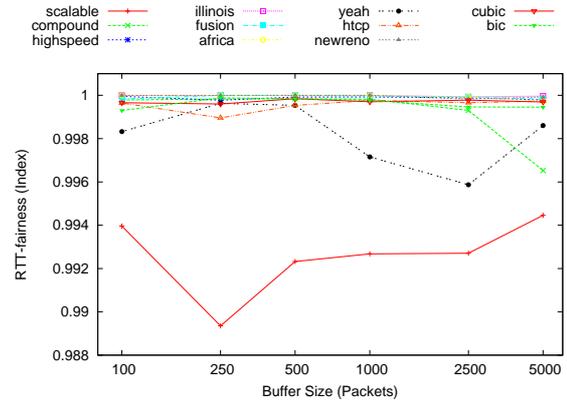}
\caption{RTT-fairness vs. Buffer Size}
\label{rttfair}
\end{figure}

As regarding to (Intra-, RTT-) fairness  as shown in Figures \ref{Intrafair} and \ref{rttfair}, Scalable, Africa, Fusion, highspeed, YeAH and NewReno are oscillating and they show different fairness index whenever the buffer size is changed. On the other hand, Compound, H-TCP, CUBIC, BIC and illinois are the best and they are mostly very close to 1 which represents the best case of fairness. As  illinois and Compound presents a poor average throughput and as CUBIC is a derived version from BIC and H-TCP so the best TCP variant in terms of average throughput, loss ratio and fairness is CUBIC followed by YeAH.

As described in Section \ref{Mot}, CUBIC is a loss-based algorithm and YeAH is a loss-delay-based algorithm. Thus, both approaches can give a higher performance than the other existing TCP variants and both can provide similar performance in the case of high-speed wired networks. Despite of all, the bandwidth utilization of CUBIC and YeAH, is still not enough to cope with the new generation of these networks.

\section{Conclusion}
\label{Conc}
In a nutshell, TCP Slow Start has a fatal problem which known as the burst loss which happens at the end of the initial stage of Slow Start. The cause of burst loss is the exponential increase of the congestion window in order to quickly utilize the bandwidth. Indeed, the burst loss can dangerously congest the bottleneck and may lead to slow down the performance of TCP and even to congestion collapse. All the aforementioned TCP variants suffer from this problem in different levels of danger.

As shown in Figure \ref{avrg-thrghpt}, it is very clear that, the smaller buffer size is the lower TCP throughput. Thus, all of the existing TCP variants still require more improvement to extend their ability to fully utilize the high-speed bandwidths, especially when the applied buffer is $near-zero$ or less than the BDP of the link. Furthermore, all the available TCP variants have preset variables which make it more static and needs a different setting for each network scenario. The existence of these preset variables makes the implementation of these TCP variants more harder and reduces its adaptation capabilities to fit different scenarios without any manual changes. Thus, in future versions of TCP protocol, the use of the preset variables should be avoided in order to increase the adaptation capabilities of the protocol.

Furthermore, CUBIC (loss-based) is the best TCP variant which overcomes all other variants in most scenarios. YeAH (loss-delay-based) shows a good performance in most cases but they (CUBIC and YeAH) still produce a huge burst loss. Consequently, this burst loss may be avoided in the future, by implementing a way which has the ability to find a safe exit point from the Slow Start phase before the occurrence of the burst loss. This safe exit point may be found by estimating the available bandwidth or by calculating the chain of the ACK arrivals.

For more details, Table \ref{comparison} shows the results of the second experiment in detail. While Throughput is the rate of successfully received packets measured as Mbps. LossRatio refers to the ratio between the total number of lost data packets to the total of sent packets. Intra-fair and RTT-fair determine whether the concurrent TCP flows are receiving a fair share of network bandwidth and time, respectively. Intra-fair and RTT-fair are measured as index from $zero$ to $one$, while the higher index is the higher fairness.

\begin{table*}[!t]
    \caption {A Performance Comparison of High-speed TCP Variants}
	\begin{center}
    \begin{tabular}{l|c c c c|c c c c}
    \hline
    & \multicolumn{4}{|c|}{100 pckts buffer}	&	\multicolumn{4}{|c}{250 pckts buffer}  \\ \hline
TCP variants & Throughput &	LossRatio &	Intra-fair & RTT-fair &	Throughput	& LossRatio	& Intra-fair & RTT-fair	\\ 			\hline
       bic 	&	453.32	&	0.11	&	0.74	&	1.00	&	752.00	&	0.09	&	1.00	&	1.00	\\ 
  compound 	&	107.89	&	0.05	&	1.00	&	1.00	&	112.30	&	0.07	&	1.00	&	1.00	\\ 
     cubic 	&	513.72	&	0.12	&	0.85	&	1.00	&	775.79	&	0.09	&	0.99	&	1.00	\\ 
 highspeed 	&	484.06	&	0.14	&	1.00	&	1.00	&	541.90	&	0.13	&	1.00	&	1.00	\\ 
      htcp 	&	149.44	&	0.07	&	0.97	&	1.00	&	490.85	&	0.09	&	0.99	&	1.00	\\ 
  illinois 	&	211.20	&	0.00	&	0.98	&	1.00	&	324.46	&	0.06	&	0.99	&	1.00	\\ 
  scalable 	&	220.54	&	0.06	&	0.98	&	0.99	&	485.10	&	0.11	&	0.98	&	0.99	\\ 
    fusion 	&	76.69	&	0.02	&	0.99	&	1.00	&	111.58	&	0.04	&	0.99	&	1.00	\\
      yeah 	&	346.04	&	0.08	&	0.82	&	1.00	&	638.62	&	0.17	&	0.98	&	1.00	\\ 
    africa 	&	315.06	&	0.08	&	0.93	&	1.00	&	343.42	&	0.09	&	0.91	&	1.00	\\ 
   newreno 	&	69.69	&	0.01	&	1.00	&	1.00	&	89.88	&	0.03	&	1.00	&	1.00	\\ 
    \hline
    \hline	
    & \multicolumn{4}{|c|}{500 pckts buffer}	&	\multicolumn{4}{|c}{1000 pckts buffer}  \\ \hline
TCP variants & Throughput &	LossRatio &	Intra-fair & RTT-fair &	Throughput	& LossRatio	& Intra-fair & RTT-fair	\\ 			\hline
       bic 	&	768.76	&	0.09	&	1.00	&	1.00	&	799.73	&	0.10	&	1.00	&	1.00	\\ 
  compound 	&	112.45	&	0.09	&	1.00	&	1.00	&	113.26	&	0.12	&	1.00	&	1.00	\\ 
     cubic 	&	780.10	&	0.10	&	1.00	&	1.00	&	800.22	&	0.10	&	1.00	&	1.00	\\ 
 highspeed 	&	601.40	&	0.10	&	1.00	&	1.00	&	670.91	&	0.12	&	0.99	&	1.00	\\ 
      htcp 	&	717.18	&	0.13	&	0.98	&	1.00	&	808.99	&	0.13	&	0.99	&	1.00	\\ 
  illinois 	&	440.99	&	0.12	&	1.00	&	1.00	&	572.63	&	0.14	&	0.99	&	1.00	\\ 
  scalable 	&	616.65	&	0.22	&	0.99	&	0.99	&	794.60	&	0.35	&	0.69	&	0.99	\\ 
    fusion 	&	149.91	&	0.08	&	0.99	&	1.00	&	241.08	&	0.10	&	0.93	&	1.00	\\ 
      yeah 	&	716.10	&	0.16	&	1.00	&	1.00	&	849.71	&	0.14	&	0.89	&	1.00	\\ 
    africa 	&	385.71	&	0.10	&	0.99	&	1.00	&	505.20	&	0.11	&	0.98	&	1.00	\\ 
   newreno 	&	120.71	&	0.05	&	1.00	&	1.00	&	173.38	&	0.05	&	0.99	&	1.00	\\
    \hline 
    \hline	
    & \multicolumn{4}{|c|}{2500 pckts buffer}	&	\multicolumn{4}{|c}{5000 pckts buffer}  \\ \hline
TCP variants & Throughput &	LossRatio &	Intra-fair & RTT-fair &	Throughput	& LossRatio	& Intra-fair & RTT-fair	\\ 			\hline
       bic 	&	891.81	&	0.12	&	0.98	&	1.00	&	958.70	&	0.12	&	0.96	&	1.00	\\ 
  compound 	&	115.56	&	0.23	&	1.00	&	1.00	&	117.38	&	0.36	&	1.00	&	1.00	\\ 
     cubic 	&	874.68	&	0.14	&	0.98	&	1.00	&	952.69	&	0.13	&	0.96	&	1.00	\\ 
 highspeed 	&	774.36	&	0.12	&	0.87	&	1.00	&	863.42	&	0.11	&	0.89	&	1.00	\\ 
      htcp 	&	854.11	&	0.14	&	0.99	&	1.00	&	920.58	&	0.15	&	1.00	&	1.00	\\ 
  illinois 	&	695.97	&	0.13	&	0.98	&	1.00	&	746.00	&	0.14	&	0.96	&	1.00	\\ 
  scalable 	&	860.40	&	0.40	&	0.83	&	0.99	&	912.83	&	0.42	&	0.73	&	0.99	\\ 
    fusion 	&	508.18	&	0.19	&	0.90	&	1.00	&	889.58	&	0.11	&	0.82	&	1.00	\\ 
      yeah 	&	890.39	&	0.18	&	0.80	&	1.00	&	966.83	&	0.16	&	0.82	&	1.00	\\ 
    africa 	&	633.71	&	0.13	&	0.91	&	1.00	&	822.56	&	0.12	&	0.80	&	1.00	\\ 
   newreno 	&	272.50	&	0.07	&	0.72	&	1.00	&	428.39	&	0.11	&	0.79	&	1.00	\\ 		\hline
    \end{tabular}
    \label{comparison}
    \end{center}
\end{table*}

\section*{Acknowledgment}
This work was supported by the Ministry of Higher Education of Malaysia under the Fundamental Research Grant FRGS/02/01/12/1143/FR for financial support.
\\





\vspace{0.25cm}
\noindent







\end{document}